\documentclass[pre,twocolumn]{revtex4-1} 
\usepackage{amsmath}  % needed for \tfrac, \bmatrix, etc.
\usepackage{amsfonts} % needed for bold Greek, Fraktur, and blackboard bold
\usepackage{graphicx} % needed for figures
\usepackage{hyperref}
\begin{document}

\title{Classical and quantum Brownian motion in an electromagnetic field~\cite{FQMT15}}
% In a long title you can use \\ to force a line break at a certain location.

\author{Marco Patriarca}
\email{marco.patriarca@kbfi.ee} % optional
%\altaffiliation[permanent address: ]{KBFI, National Institute of Chemical Physics and Biophysics, R\"avala 10, Tallinn 15042, Estonia} % optional second address
% If there were a second author at the same address, we would put another 
% \author{} statement here.  Don't combine multiple authors in a
% single \author statement.
\affiliation{NICPB--National Institute of Chemical Physics and Biophysics, R\"avala 10, Tallinn 15042, Estonia}

\author{Pasquale Sodano}
%\email{pasquale.sodano02@gmail.com} % optional
%\altaffiliation[permanent address: ]{KBFI, National Institute of Chemical Physics and Biophysics, R\"avala 10, Tallinn 15042, Estonia} % optional second address
% If there were a second author at the same address, we would put another 
% \author{} statement here.  Don't combine multiple authors in a
% single \author statement.
\affiliation{International Institute of Physics, Universidade Federal do Rio Grande do Norte, Natal-RN 59078-400, Brazil}

% See the REVTeX documentation for more examples of author and affiliation lists.

\date{\today}

\begin{abstract}
The dynamics of a Brownian particle in a constant magnetic field and time-dependent electric field is studied in the limit of white noise, using a Langevin approach for the classical problem and the path-integral Feynman-Vernon and Caldeira-Leggett framework for the quantum problem.
We study the time evolution in configuration space of the probability distribution of an initial pure state represented by an asymmetrical Gaussian wave function and show that it can be described as the superposition of (a) the classical motion of the center of mass, (b) a rotation around the mean position, and (c) a spreading processes along the principal axes.
\end{abstract}

\maketitle

%\newpage

% ==================================================================================
\section{Introduction}
% ==================================================================================
The problem of a Brownian particle in a magnetic field arises in different fields from condensed-matter (e.g. the Hall effect) to cosmology (e.g. cosmic rays).
The present paper focuses on Brownian motion in a constant magnetic field and a spatially homogeneous, possibly time-dependent) electromagnetic field, a problem studied so far by various authors both in the classical regime
\cite{Taylor1961a,Kursunoglu1962a,Kursunoglu1963a,Liboff1966a,Karmeshu1973a,Karmeshu1974a,Xiang1993a,Singh1996a,Lemons1999a,Czopnik2001a,Czopnik2001b,Dodin2005a,Simoes2005a,Jimenez2006a,Jimenez2007a,Jimenez2008c,Paraan2008a,Voropajeva2008a,Hou2009a,Dattagupta2010a,Lagos2011a,Dattagupta2014a,Friz2015a}
and in the quantum regime
\cite{Furuse1970a,Das1981a,Das1982a,Jayannavar1981a,Marathe1989a,Dattagupta1996a,Mitra2010a,
Xiang1993a,Dattagupta2010a,Dattagupta2014a}.

In the absence of noise and dissipation, a close analogy links a particle in a constant magnetic field to the harmonic oscillator, both in the classical \cite{Landau2} and in the quantum \cite{Landau3} problem.
At a classical level, the particle performs periodic harmonic motion with frequency $\omega$, with $\omega = \sqrt{k/m}$ for a harmonic oscillator of mass $m$ and elastic constant $k$ and with the cyclotron frequency $\omega = |q\boldsymbol{B}|/mc$ for a particle of charge $q$ in a magnetic field $\boldsymbol{B}$ ($c$ is the speed of light).
At a quantum level, the energy spectrum of both systems has equidistant energy levels $E_n$ with energy spacing $\Delta E = E_{n+1} - E_n = \hbar\omega$.
However, an arbitrary small internal noise breaks this analogy and turns the harmonic-like motion of a particle in a magnetic field into one similar to that of a free Brownian particle, see Refs. \cite{Taylor1961a,Kursunoglu1962a,Singh1996a,Karmeshu1974a,Czopnik2001a} for the classical case and Refs. \cite{Marathe1989a,Li1990a,Dattagupta1996a} for the quantum case.
%This basic difference --- underlying the apparent similarity --- consistent with the Bohr-van Leeuwen theorem \cite{Bohr1911a,Leeuwen1921a}, stating that the state of a system at thermal equilibrium state cannot be affected by the presence of a vector potential.
In this paper we study the time evolution of the probability density in configuration space of a Brownian particle in a constant magnetic field and a homogeneous electric field, in the white noise approximation.
We outline the main steps in the derivation of the results --- further details will be presented elsewhere.
For the quantum problem, the time evolution of an asymmetrical Gaussian wave packet is worked out within the framework of the Feynman-Vernon \cite{Feynman1963a} and Caldeira-Leggett \cite{Caldeira1983a} models.

% ==================================================================================
\section{Classical problem}
\label{classical}
% ==================================================================================

In the Langevin approach a classical Brownian particle in an electromagnetic field is described (in the limit of white noise) by the following stochastic equation \cite{Singh1996a},
\begin{equation}
  m \frac{d\boldsymbol{v}}{dt} = q\boldsymbol{E} + \frac{q}{c} \boldsymbol{v} \times \boldsymbol{B} - m \gamma \boldsymbol{v} + \boldsymbol{R}(t)\ .
  \label{EM}
\end{equation}
Here $\boldsymbol{v}(t) = (v_x(t),v_y(t),v_z(t)) = d\boldsymbol{r}(t)/dt$ is the particle velocity at time $t$;
$\boldsymbol{r}(t) = (x(t),y(t),z(t))$ is the particle position;
the first two terms on the right hand side are the electric and the Lorentz force, respectively;
the last two terms represent the environment forces, i.e., the dissipative force $-m\gamma\boldsymbol{v}$ ($\gamma$ is the friction coefficient) and the random force $\boldsymbol{R}(t) = (R_x(t),R_y(t),R_z(t))$, assumed as a Gaussian zero-mean  $\delta$-correlated stochastic process ($i,j = x,y,z,$),
\begin{eqnarray}
  \left\langle R_i(t) \right\rangle = 0\ ,
  ~~~
  \left\langle R_i(t) R_j(s) \right\rangle = \frac{2m\gamma}{\beta}\delta(t-s)\delta_{ij},
  \label{RR}
\end{eqnarray}
where $\beta = 1/k_B T$ is the inverse temperature and $\left\langle \dots \right\rangle$ represents a statistical averaging over the stochastic force configurations.

Decomposing the velocity $\boldsymbol{v} = \boldsymbol{v}_\parallel + \boldsymbol{v}_\perp$, with $\boldsymbol{v}_\parallel$ parallel and $\boldsymbol{v}_\perp$ perpendicular to the magnetic field, and analogously for the electric field, $\boldsymbol{E} = \boldsymbol{E}_\parallel + \boldsymbol{E}_\perp$, and the random force, $\boldsymbol{R} = \boldsymbol{R}_\parallel + \boldsymbol{R}_\perp$, Eq.~(\ref{EM})becomes decoupled,
\begin{eqnarray}
    m \frac{d\boldsymbol{v}_\parallel}{dt} 
  & = & q\boldsymbol{E}_\parallel - m \gamma \boldsymbol{v}_\parallel + \boldsymbol{R}_\parallel(t) \, ,
  \label{par}
  \\
  m \frac{d\boldsymbol{v}_\perp}{dt} 
  & = & q\boldsymbol{E}_\perp + \frac{q}{c} \boldsymbol{v}_\perp \times \boldsymbol{B} - m \gamma \boldsymbol{v}_\perp + \boldsymbol{R}_\perp(t) \, .
  \label{per}  
\end{eqnarray}
%
%%%%%%%%%%%%%%%%%%%%%%%%%%%%%%%%%%%%%%%%%%%%%%
\begin{figure}[ht!]
\centering
\includegraphics[width=6cm]{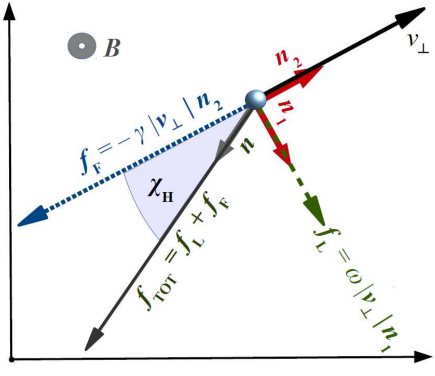}
\caption{\label{forces}
Lorentz force $f_\mathrm{L}$, friction force $f_\mathrm{F}$, and resulting generalized friction force $f_\mathrm{TOT}$ proportional to $|\boldsymbol{v}_\perp|$, forming a constant angle $\chi_\mathrm{H}$ with respect to $\boldsymbol{v}_\perp$; see text for details.}
\end{figure}
%%%%%%%%%%%%%%%%%%%%%%%%%%%%%%%%%%%%%%%%%%%%%%%
The motion parallel to the magnetic field is equivalent to that of a one-dimensional Langevin particle acted upon only by the force $q\boldsymbol{E}_\parallel(t)$ and will not be considered further.
%As for the motion in the plane perpendicular to the magnetic field,
It is now useful to define the versors
\begin{eqnarray}
\boldsymbol{n}_2 = \frac{\boldsymbol{v}_\perp^{\,}}{|\boldsymbol{v}_\perp|^{\,}} \, ,
~~~~
\boldsymbol{n}_1 = \boldsymbol{n}_2 \times \frac{\boldsymbol{B}}{|\boldsymbol{B}|} \, ,
\end{eqnarray}
see Fig. \ref{forces}, and rewrite Eq.~(\ref{per}) as
\begin{eqnarray}
  &&\frac{d\boldsymbol{v}_\perp}{dt} 
  = \frac{q}{m}\boldsymbol{E}_\perp
      + ( \omega \, \boldsymbol{n}_1 - \gamma \, \boldsymbol{n}_2 )  |\boldsymbol{v}_\perp|
      + \frac{1}{m} \boldsymbol{R}_\perp(t),~~
\nonumber \\
  &&= \frac{q}{m}\boldsymbol{E}_\perp
      + (\sin\chi_\mathrm{H}\,\boldsymbol{n}_1 \!-\! \cos\chi_\mathrm{H}\,\boldsymbol{n}_2) \varGamma |\boldsymbol{v}_\perp|
      + \frac{1}{m} \boldsymbol{R}_\perp(t).~~
  \label{per2}  
\end{eqnarray}
Here $\sin\chi_\mathrm{H} = \omega/\varGamma$, $\cos\chi_\mathrm{H} = \gamma/\varGamma$ are the force direction cosines in the $(\boldsymbol{n}_1,\boldsymbol{n}_2)$ frame and 
\begin{eqnarray}
  \label{Hall-Gamma}
  \tan\chi_\mathrm{H} = \frac{\omega}{\gamma} \ ,
~~~~~
  \varGamma  = \sqrt{\gamma^2 + \omega^2} \ ,
\end{eqnarray}
are the Hall angle \cite{Ashcroft1976a} and the ``effective damping constant''.
Equation (\ref{per2}) displays the similarity between the Lorentz and the friction forces, both proportional to $|\boldsymbol{v}_\perp|$, merging in an effective viscous force still proportional to $|\boldsymbol{v}_\perp|$ with intensity $\varGamma$, forming an angle $\chi_\mathrm{H}$ (the Hall angle) with $\boldsymbol{v}_\perp$ --- see Fig. \ref{forces}. 
Note that even if $\boldsymbol{n}_1(t)$ and $\boldsymbol{n}_2(t)$ are time-dependent, $\varGamma$ and $\chi_\mathrm{H}$ are constant in time.
The Hall angle $\chi_\mathrm{H}$ (known from the Hall effect \cite{Ashcroft1976a} where both a magnetic and an electric field are present) measures the relative strengths of Lorentz to friction force, playing a key role also when only a magnetic field is present.

The time scales associated to $\omega$ and $\gamma$ and the parameters $\varGamma$ and $\chi_\mathrm{H}$ can be related to each other through the \emph{complex friction coefficient}
\begin{equation}
  \widetilde{\varGamma} = \gamma + i \omega = \varGamma \exp \left( i\chi_\mathrm{H} \right) \ ,
  \label{CGamma}
\end{equation}
which is relevant both in the classical and in the quantum problem --- $\gamma$ and $\omega$ represent the real and imaginary parts of $\widetilde{\varGamma}$, while the Hall angle $\chi_\mathrm{H}$ and the effective damping constant $\varGamma$ represent its modulus and phase, respectively.

% ------------------------------------------------------------------- 
\subsection{Constant $\boldsymbol{B}$ and homogeneous $\boldsymbol{E}(t)$}
% -------------------------------------------------------------------
%
We assume a homogeneous electric field $\boldsymbol{E}(t)$, an $\omega = qB/mc > 0$, and a constant magnetic field,
\begin{eqnarray}
   \label{B} \boldsymbol{B} = (0,0,B) \, .
\end{eqnarray}
It is convenient to introduce the complex coordinate \cite{Landau2}
\begin{equation}
  Z(t) = x(t) + i y(t)\ ,
  \label{Zeta}
\end{equation}
and the complex forces $q\widetilde{E}(t) = qE_x(t) + i qE_y(t)$, $\widetilde{R}(t) = R_x(t) + i R_y(t)$.
Using Eq.~(\ref{EM}), one finds that $Z(t)$ follows the complex Langevin equation
\begin{equation}
  m\ddot{Z}(t) = - m\widetilde{\varGamma}\dot{Z}(t) + q\widetilde{E}(t) + \widetilde{R}(t) \, .
\label{ZEM}
\end{equation}
Here the Lorentz and friction forces merged in the generalized friction force given by the term
$- m \widetilde{\varGamma} \dot{Z}(t)$,
where $\widetilde{\varGamma}$ is defined in Eqs.~(\ref{Hall-Gamma}).
By integrations of Eq.~(\ref{ZEM}) between $t = t_a$ and $t = t_b$, one obtains the complex velocity $\dot{Z}(t_b)$ and coordinate $Z(t_b)$,
\begin{eqnarray}
  & \dot{Z}(t_b) & = \dot{Z}_0 \exp\left(-\widetilde{\varGamma}t\right) 
                 + \frac{1}{m} \dot{F}[q\widetilde{E} + \widetilde{R}\,],~~
  \label{Vsolution}
  \\
  & Z(t_b) & = Z_0 
           + \frac{\dot{Z}_0}{\widetilde{\varGamma}} \left[ 1-\exp\left(-\widetilde{\varGamma}t\right) \right]
           + \frac{1}{m} \! F[q\widetilde{E} + \widetilde{R}].~~
  \label{solution}
\end{eqnarray}
Here $t = t_b - t_a$, $Z_0 = Z(t_a) = x(t_a) + iy(t_a)$, $\dot{Z}_0 = \dot{Z}(t_a) = \dot{x}(t_a) + i\dot{y}(t_a)$;
the functionals $F$ and $\dot{F} = dF[f]/dt_b$ represent the inhomogeneous contributions of $\widetilde{E}$ and $\widetilde{R}$,
\begin{eqnarray}
\label{F}
F[\tilde{f}\,] = \frac{1}{\widetilde{\varGamma}} 
                 \int_{t_a}^{t_b}ds\,
                  \left\{ 1 - \exp\left[ -\widetilde{\varGamma}(t_b - s)\right] \right\} \tilde{f}(s) ,
\end{eqnarray}
where $\tilde{f} \equiv q\widetilde{E} + \widetilde{R}$.

% ----------------------------------------------------------------------
\subsection{Constant magnetic field}
% ----------------------------------------------------------------------
%
Here we consider the particular case of zero electric field.
For convenience we rewrite the initial conditions on the complex variable $Z$ as
\begin{eqnarray}
   Z_0 & = x_0 + i y_0 \ ,
   ~~~~~~
   \dot{Z}_0 & = \dot{x}_0 + i\dot{y}_0 \equiv v_0 \exp\left( i\varphi_0 \right),~~
  \label{z0}
\end{eqnarray}
where $x_0 \equiv x(t_a)$, $\dot{x}_0 \equiv \dot{x}(t_a)$, and analogously for the $y$ components, and we have introduced the initial modulus $v_0 = |\boldsymbol{v}_0|$ and angle $\varphi_0$ formed with the $x$-axis,
\begin{eqnarray}
  v_0 = \sqrt{\dot{x}_0^2+\dot{y}_0^2}  ,
  ~~~~~
  \tan\varphi_0 ~ = \frac{ \dot{y}_0 }{ \dot{x}_0 } \, .
  \label{v0}
\end{eqnarray}
From Eqs.~(\ref{solution}) and (\ref{RR}), separating $\left\langle Z(t) \right\rangle$ into real and imaginary parts, one obtains the average coordinates,
\begin{eqnarray}
  \left\langle x(t) \right\rangle &=& x_c - 
          \frac{v_0}{\varGamma}\exp(-\gamma t) \cos\left(\omega t-\varphi_0+\chi_\mathrm{H}\right)\ ,
  \nonumber\\
  \left\langle y(t) \right\rangle &=& y_c + 
          \frac{v_0}{\varGamma}\exp(-\gamma t) \sin\left(\omega t-\varphi_0+\chi_\mathrm{H}\right)\ ,
  \label{xy-E0}
\end{eqnarray}
where $t=t_b-t_a$ and $x_c$, $y_c$ are the coordinates of the position eventually approached for $t \gg \gamma^{-1}$,
\begin{eqnarray}
  x_c &=& \langle x(t \! \gg \! \gamma^{-1}\!) \rangle = x_0 + \frac{v_0}{\varGamma}\cos(\chi_\mathrm{H}-\varphi_0)\ ,
  \nonumber\\
  y_c &=& \langle y(t \! \gg \! \gamma^{-1}\!) \rangle  = y_0 - \frac{v_0}{\varGamma}\sin(\chi_\mathrm{H}-\varphi_0)\ .
\label{xcyc}
\end{eqnarray}
This provides a simple geometrical interpretation of the Hall angle $\chi_\mathrm{H}$ as the total deflection angle of the particle with respect to its initial velocity, while the effective friction coefficient $\varGamma$ defines the total distance covered by the particle, $d_B = [(y_c - y_0)^2 + (x_c - x_0)^2 ]^{1/2} = v_0/\varGamma$.

In polar coordinates, choosing the origin in the asymptotic position $(x_c,y_c)$, from Eqs.~(\ref{xy-E0}) one obtains
\begin{eqnarray}
  & r(t) & = \sqrt{[\left\langle x(t) \right\rangle - x_c]^2 + [\left\langle y(t) \right\rangle - y_c]^2} 
       = \frac{v_0}{\varGamma} \, \mathrm{e}^{-\gamma t} ,
  \label{r}
  \\
  & \theta(t) & = \arctan\left[ \frac{\left\langle y(t) \right\rangle - y_c}{\left\langle x(t) \right\rangle - x_c } \right] 
            = \varphi_0 - \chi_\mathrm{H} - \omega t\, .
  \label{theta_c}
\end{eqnarray}
By eliminating the time variable one finds that the shape of the trajectory is an exponential spiral,
\begin{equation}
  r(\theta) = 
  \frac{v_0}{\varGamma} \exp\left( -\frac{ \theta - \varphi_0 + \chi_\mathrm{H} } { \tan\chi_\mathrm{H} } \right)\, .
  \label{spiral}
\end{equation}
Notice in Eq.~(\ref{r}) that the particle approaches the asymptotic position $(x_c,y_c)$ with time scale $\gamma^{-1}$ --- as in the problem without magnetic field --- and, at the same time, Eq.~(\ref{theta_c}) shows a \emph{uniform} angular motion with angular velocity $\omega$ --- as in the frictionless problem with a constant magnetic field.
These complementary features are due to the fact that the Lorentz force acts perpendicularly to the particle velocity, leaving its modulus and therefore the relaxation dynamics unaffected, so that the only effect of the magnetic field is to bend the trajectory. Whereas the friction force --- being anti-parallel to the particle velocity--- changes the velocity modulus as if no magnetic field is present, without influencing the direction.

The position uncertainties can be expressed through the second moments
\begin{eqnarray}
& \left\langle\Delta x(t)^2\right\rangle & = \left\langle[x(t)-\left\langle x(t)\right\rangle]^2\right\rangle \, ,\\
& \left\langle\Delta y(t)^2\right\rangle & = \left\langle [y(t)-\left\langle y(t)\right\rangle]^2\right\rangle \, , \\
& \left\langle\Delta x(t) \Delta y(t)^{\!} \right\rangle & =
  \left\langle [x(t) \!-\! \left\langle x(t)\right\rangle] \!\times\! [y(t) \!-\! \left\langle y(t)\right\rangle]^{}\right\rangle \, ,
\end{eqnarray}
where $\Delta x(t) = x(t)-\left\langle x(t)\right\rangle$ and $\Delta y(t) = y(t)-\left\langle y(t)\right\rangle$ are the $x$ and $y$ displacements, which can be computed from $\langle \Delta Z(t)^2 \rangle$ and $\langle|\Delta Z(t)|^2 \rangle$.
One finds $\langle\Delta x(t)^2\rangle = \langle\Delta y(t)^2\rangle$ and a radial mean square displacement \cite{Singh1996a,Williamson1968a}
\begin{eqnarray}
  && \langle\Delta r(t)^2\rangle
  = \langle\Delta x(t)^2\rangle + \langle\Delta y(t)^2\rangle
  = \frac{2 \cos\!\chi_\mathrm{H}^{\ 2}}{m\beta\gamma^2} 
      \bigg\{ \, 2 \gamma t~~~
        \nonumber \\
  &&    + 4 \cos\chi_\mathrm{H}
            \Big[
            \mathrm{e}^{-\gamma t} \!\cos(\omega t \!+\! \chi_\mathrm{H})
            \!-\!  \cos\chi_\mathrm{H}
            \Big]
        \!+\! 1
        \!-\! \mathrm{e}^{-2\gamma t} 
       \!\bigg\}.~~~
  \label{dr}
\end{eqnarray}
For $\chi_\mathrm{H} \to 0$ ($B \to 0$) this expression reduces to the mean square displacement of a Langevin particle in a plane \cite{Wang1945a}, while in the asymptotic limit $\gamma t \gg 1$ one obtains
\begin{equation}
  \left\langle\Delta r^2\right\rangle_t = 4 D_B t \ ,
   ~~~~~
   D_B = \frac{\gamma}{m\beta\varGamma^2} \equiv {\rm Re}\left(\frac{1}{m\beta\widetilde{\varGamma}}\right) \ .
  \label{DB}
\end{equation}
The latter equation resembles the Einstein relation, with $1/\gamma$ replaced by (the real part of) $1/\widetilde{\varGamma}$.

% -------------------------------------------------------------------------------------
\subsection{Constant electromagnetic field}
% -------------------------------------------------------------------------------------
%
It is easy to show from Eq.~(\ref{solution}) that a homogeneous electric field modifies the mean position but not the mean square displacements.
The well known case of a constant electric field $\boldsymbol{E}$ with modulus $E_0$ and components $E_x=E_0\cos\alpha$ and $E_y=E_0\sin\alpha$ can be recovered from Eq.~(\ref{xy-E0}) and is characterized by an asymptotic motion with an angle $\chi_\mathrm{H}$ with respect to the electric field,
\begin{eqnarray}
  && \left\langle x(t)\right\rangle  = \left\langle x(t)\right\rangle_{\boldsymbol{E}=0} + qE_0\cos(\alpha-2\chi_\mathrm{H})/m\varGamma^2 ,
  \nonumber\\
  &&~~~~~              + (qE_0)/(m\varGamma^2)\exp(-\gamma t)\cos\left(\omega t+2\chi_\mathrm{H}-\alpha\right) + V_x t \ ,
  \nonumber\\
  && \left\langle y(t)\right\rangle  = \left\langle y(t)\right\rangle_{\boldsymbol{E}=0} + qE_0\sin(\alpha-2\chi_\mathrm{H})/m\varGamma^2
  \nonumber\\
  &&~~~~~              - (qE_0)/(m\varGamma^2)\exp(-\gamma t)\sin\left(\omega t+2\chi_\mathrm{H}-\alpha\right) + V_y t \ ,
  \label{xy-E}
\end{eqnarray}
where the drift velocities $V_x t$ and $V_y t$ are given by
\begin{eqnarray}
  V_x = \frac{qE_0}{m\varGamma}\cos(\alpha-\chi_\mathrm{H}) \ ,
  ~~~~~~
  V_y = \frac{qE_0}{m\varGamma}\sin(\alpha-\chi_\mathrm{H}) \ .
  \label{Vxy}
\end{eqnarray}
The velocity modulus is $V_0 = qE_0/m\varGamma$, confirming the role of $\varGamma$ as an effective friction coefficient.

% ==============================================================================
\section{Quantum problem}
\label{quantum}
% ==============================================================================

In this section we study the time evolution of a quantum wave packet between times $t_a$ and $t_b$ ($t_a < t_{\rm b}$).

% --------------------------------------------------------------------------------
\subsection{Initial state}
\label{initial_state}
% --------------------------------------------------------------------------------

It is assumed that the wave function $\varPsi(x_a,y_a,z_a,t_a)$ at the initial time $t_a$ is known and factorized,
\begin{equation}
  \varPsi(x_a,y_a,z_a,t_a) = \psi(x_a,y_a,t_a) \times \phi(z_a,t_a)  \ ,
  \label{Psi_a}
\end{equation}
so that the $x$-$y$ motion, on which we concentrate, is decoupled from the $z$-motion.
The problem is undetermined by a phase factor due to gauge invariance, i.e., if $\psi \to \psi \times \exp[i\varLambda]$ ($\varLambda \equiv \varLambda(x,y)$ is an arbitrary function) and ${\bf A} \to {\bf A} + \nabla\varLambda$.
We assume a magnetic field ${\bf B}=(0,0,B)\equiv\nabla\times{\bf A}$ coming from a vector potential
\begin{equation}
  {\bf A} = ( \ (\lambda-1)By/2, \ (\lambda+1)Bx/2, \ 0 \ ) \ ,
  \label{A}
\end{equation}
with arbitrary gauge parameter $\lambda$.
A Gaussian shape is assumed for the initial wave function in the $x$-$y$-plane,
\begin{eqnarray}
  &&\psi(x_a,y_a,t_a)
    = \sqrt{N_0} \exp\bigg[
     -\frac{(x_a-x_0)^2}{4\sigma_x^2}
     -\frac{(y_a-y_0)^2}{4\sigma_y^2} 
  \nonumber \\
  &&+ \frac{i}{\hbar}\left( q_x x_a + q_y y_a + \frac{\mu m \omega}{2} x_a y _a \right) 
     \bigg] \, ,
  \label{psi_a}
\end{eqnarray}
associated to a Gaussian probability density,
\begin{eqnarray}
  &&P(x_a,y_a,t_a) = |\psi(x_a,y_a,t_a)|^2 \nonumber\\
  &&= N_0 \exp\left[ - \frac{(x_a-x_0)^2}{2\sigma_x^2} - \frac{(y_a-y_0)^2}{2\sigma_y^2} \right] .
  \label{P_a}
\end{eqnarray}
Here $N_0=1/2\pi\sigma_x\sigma_y$ is a normalization factor, the coordinates $(x_0,y_0) = (\left\langle x_a \right\rangle, \left\langle y_a \right\rangle)$ are the initial average positions ($\left\langle\dots\right\rangle$ denoting in this section a quantum average), and $\sigma_x$, $\sigma_y$ are the corresponding standard deviations, $\left\langle (x_a-x_0)^2 \right\rangle = \sigma_x^2$, $\left\langle (y_a-y_0)^2 \right\rangle = \sigma_y^2$.
We assumed, without loss of generality, that there is no mixed term $\propto x y$ and an anisotropic wave function with $\sigma_x \ne \sigma_y$.
Finally, the parameters $q_x$, $q_y$ and $\mu$ are related to the average initial velocity $\widehat{{\bf v}} = (v_x,v_y) = \widehat{{\bf p}}/m = - i\nabla/m\hbar - e{\bf A}/mc$, that from Eq. (\ref{psi_a}) is
\begin{eqnarray}
  v_x = \frac{q_x}{m} + \frac{\mu \!-\! \lambda \!+\! 1}{2} \omega y_0,
    ~
  v_y = \frac{q_y}{m} + \frac{\mu \!-\! \lambda \!-\! 1}{2} \omega x_0.~~~
  \label{vx_vy}
\end{eqnarray}
Note that $\widehat{{\bf v}}$ (as any other observable quantity) only depends on the difference
\begin{equation}
  \alpha = \mu - \lambda \ ,
  \label{alpha}
\end{equation}
exhibiting the gauge invariance of the problem.

The quantum treatment discussed below is based on the density matrix, given at the initial time $t=t_a$ by
\begin{equation}
  \rho(x_a,y_a,x_a',y_a',t_a) = \psi(x_a,y_a,t_a) \psi^*(x_a',y_a',t_a) \ .
  \label{rho_a}
\end{equation}
It is convenient to introduce the new coordinates
\begin{eqnarray}
  X = \frac{x' \!+\! x}{2},
  ~
  Y = \frac{y' \!+\! y}{2},
  ~
  \xi = \frac{x' \!-\! x}{2},
  ~
  \eta = \frac{y' \!-\! y}{2},~~
  \label{X_Y_xi_eta}
\end{eqnarray}
with inverse relations $x = X - \xi/2$, $y = Y - \eta/2$, and $x' = X + \xi/2$, $y' = Y + \eta/2$, that are useful for simplifying calculations and because they provide a physical interpretation of the results.
In terms of the new variables, the initial density matrix becomes
\begin{eqnarray}
  &&\rho(X_a,Y_a,\xi_a,\eta_a,t_a) 
  \nonumber \\
  &=& \psi\left( X_a-\frac{\xi_a}{2},Y_a-\frac{\eta_a}{2} \right)
      \psi^*\left( X_a+\frac{\xi_a}{2},Y_a+\frac{\eta_a}{2} \right)
  \nonumber \\
  &=& N_0 \exp\bigg\{ -\frac{(X_a-x_0)^2}{2\sigma_x^2}
                     -\frac{(Y_a-y_0)^2}{2\sigma_y^2} 
                     -\frac{\xi_a^2}{8\sigma_x^2}
                     -\frac{\eta_a^2}{8\sigma_y^2} 
  \nonumber \\
       &&  - \frac{i}{\hbar}\left( q_x \xi_a 
                                    +q_y \eta_a 
                                    +\frac{\mu m\omega}{2}\left(\xi_a Y _a+X_a\eta_a\right)
                                  \right) 
               \bigg\} \ .
  \label{Rho_a}
\end{eqnarray}
%

% --------------------------------------------------------------------------------
\subsection{Time evolution}
\label{time_evolution}
% --------------------------------------------------------------------------------

The reduced density matrix $\rho(X,\-Y,\-\xi,\-\eta,t)$ evolves with time in a way similar to the wave function,
\begin{eqnarray}
  &&\rho(X_b,Y_b,\xi_b,\eta_b,t_b) \\
  &&= \int dX_a dY_a d\xi_a d\eta_a J(x_b,y_b,t_b|x_a,y_a,t_a) \rho(x_a,y_a,t_a) \, ,\nonumber
  \label{rho_b}
\end{eqnarray}
where the effective propagator $J$ is conveniently expressed in terms of the effective action $S[X,Y,\xi,\eta]$ as
\begin{eqnarray}
  &&J(X_b,Y_b,\xi_b,\eta_b,t_b|X_a,Y_a,\xi_a,\eta_a,t_a) 
  \nonumber \\
  &&= \int_a^{\,b} \!\! DX DY D\xi D\eta
               \exp\left\{\frac{i}{\hbar} S[X,Y,\xi,\eta]\right\}  .
  \label{J_ba}
\end{eqnarray}
Here $a$ represents the boundary conditions at time $t_a$: $X(t_a)=X_a$, $Y(t_a)=Y_a$, $\xi(t_a)=\xi_a$, and
$\eta(t_a)=\eta_a$, and analogously for $b$.
For an isolated system the effective action $S[X,Y,\xi,\eta]$ is the difference between the actions of the isolated system,
$S[X,Y,\xi,\eta] \!=\! S_0[X \!-\! {1 \over 2} \xi, \!Y \!-\! {1 \over 2} \eta] \!-\! S_0[X \!+\! {1 \over 2} \xi, \!Y \!+\! {1 \over 2}\eta]$
--- and the effective propagator factorizes in the product of two propagators for $\psi(x,y,t)$ and $\psi^*(x',y',t)$, where, for a non-relativistic particle \cite{Landau3},
\begin{equation}
  S_0[x,y] = \int_{t_a}^{t_b} dt \left[ \frac{m}{2}\left( \dot{x}^2 + \dot{y}^2 \right)
                                 + \frac{e}{c}\left( \dot{x}A_x + \dot{y}A_y \right) \right].
  \label{S_0}
\end{equation}
For a non-isolated system, the effective action contains also the \emph{influence phase} $\varPhi_2$, coming from the integration of the environment degrees of freedom~\cite{Feynman1963a,Feynman1965a},
\begin{eqnarray}
  S[X,Y,\xi,\eta]
  &=& S_0\left[ X-\frac{\xi}{2},Y-\frac{\eta}{2}\right] 
    - S_0\left[ X+\frac{\xi}{2},Y+\frac{\eta}{2}\right]
    \nonumber
    \\
  &+& \hbar \, \varPhi_2[X,Y,\xi,\eta] \ .
  \label{S_eff}
\end{eqnarray}
Here we focus on the dynamics of the probability density $P(X,Y,t)$ in configuration space, which is given by the diagonal elements of $\rho(x,x',y,y',t)$, i.e., $P(X,Y,t) = \left[\rho(x,x',y,y',t)\right]_{x=x'=X; \, y=y'=Y}$.
If the coordinates $X,Y,\xi,\eta$ are used, one can obtain $P$ setting $\xi=\eta=0$, i.e., $P(X,Y,t)=\rho(X,Y,\xi=0,\eta=0,t)$.
Then the probability density at time $t_b$, from Eqs.~(\ref{J_ba}), is
\begin{eqnarray}
  &&P(X_b,Y_b,t_b)
  = \rho(X_b,Y_b,0,0,t_b)
  \nonumber
  \\
  &&= \int dX_a dY_a d\xi_a d\eta_a
  J(X_b,Y_b,0,0,t_b|X_a,Y_a,\xi_a,\eta_a,t_a)\times
  \nonumber
  \\
  &&\times\rho(X_a,Y_a,\xi_a,\eta_a,t_a) \ .
  \label{P_b}
\end{eqnarray}
We now proceed to obtain $\varPhi_2[X,Y,\xi,\eta]$ and $J$.

% --------------------------------------------------------------------------------
\subsection{Influence functional and phase}
\label{Phi_2}
% --------------------------------------------------------------------------------

The influence phase $\varPhi$ --- or equivalently the influence functional $\mathcal{F}=\exp(i\varPhi)$ --- is the central quantity in the study of a non-isolated quantum system, since it describes the effect of the interaction with the environment.
In the framework of the Feynman-Vernon model \cite{Feynman1963a} a dissipative environment is represented in one dimension 
by an infinite set of harmonic oscillators interacting with the central particle.
We make the simple hypothesis that the two-dimensional environment is obtained by a straightforward generalization of the one-dimensional case assuming that the central particle interacts with a set of two-dimensional
oscillators, which are harmonic and isotropic in the $x$-$y$-plane.

Thus, the starting action of the total system, depending on the coordinates of the central particle $(x,y)$ and of the oscillators $\{x_n,y_n\}$ ($n=1,2,\dots$) \cite{Patriarca1996a} is
\begin{eqnarray}
  &&S_{\rm tot}[x,y,\{x_n\},\{y_n\}]
  = S_0[x,y]
  \nonumber \\
  &&  \!+ \, \frac{m}{2} \!\! \sum_n \!\! \int_{t_a}^{t_b} \!\!\!\!\! dt 
        \big\{ \dot{x}_n^2 \!+\! \dot{y}_n^2
               \!+\! \omega_n^2 \left[ (x_n \!-\! x)^2 \!+\! (y_n \!-\! y)^2 \right]\!\!
        \big\}.
  \label{S}
\end{eqnarray}
Notice that the oscillators have been assigned their equilibrium positions coinciding with the central particle position, as requested by general symmetry constraints \cite{Patriarca1996a}.
This determines both the structure of the total Lagrangian, Eq. (\ref{S}), and the form of the (correlated) thermal equilibrium initial conditions --- see Ref. \cite{Patriarca1996a} for details.
In Eq.~(\ref{S}), the interaction between central particle and environment oscillators factorizes into a sum of two terms for the $x$ and the $y$ dimension.
Correspondingly, $x$- and $y$-factorized thermal equilibrium initial conditions at time $t=t_a$ are assumed for the $x$ and $y$ coordinates of the environment oscillators in the form of (one-dimensional) equilibrium density matrices at temperature $\beta^{-1}$.
Within the Caldeira-Leggett model \cite{Caldeira1983a}, the integration of the $x$ or $y$ environment degrees of freedom proceeds as in the 1D case \cite{Patriarca1996a} and results in the influence phase
\begin{equation}
  \varPhi_2[X,Y,\xi,\eta] = \varPhi_1[X,\xi] + \varPhi_1[Y,\eta] \ ,
  \label{Phi_fact}
\end{equation}
(and $\mathcal{F}_2[X,Y,\xi,\eta] = \mathcal{F}_1[X,\xi] \mathcal{F}_1[Y,\eta]$), implying as a limit of the present model the hypothesis of dynamical and statistical independence of the $x$ and $y$ environment degrees of freedom (second properties of influence functionals \cite{Feynman1963a}).
In the white noise approximation the 1D influence phase reads
\begin{equation}
  \hbar\varPhi_1[X,\xi] 
  = m \gamma \int_{t_a}^{t_b} dt \left[\xi(t) X(t) + \frac{i}{\beta\hbar}\xi^2(t) \right] \ .
  \label{Phi_1}
\end{equation}
Replacing $\varPhi_1$ and the unperturbed action~(\ref{S_0}) in Eq.~(\ref{S_eff}) one obtains the effective action 
\begin{eqnarray}
  &&S[X,Y,\xi,\eta]
    = m \int_{t_a}^{t_b} dt 
    \bigg[\! - \dot{X}\dot{\xi} - \dot{Y}\dot{\eta}
  \nonumber \\
  &&+ \frac{\omega(\dot{X}\eta \!+\! Y\dot{\xi} \!-\! X\dot{\eta} \!-\! \dot{Y}\xi)}{2} 
        \!+\! \gamma(\dot{X}\xi \!+\! \dot{Y}\eta)
        \!+\! \frac{i\gamma(\xi^2 \!+\! \eta^2)}{\beta\hbar} 
    \!\bigg].~~~~~
  \label{S_eff1}
\end{eqnarray}
The first two terms in the integral are the kinetic ones, the term $\propto\omega$ is due to the magnetic field, the real term $\propto\gamma$ to dissipation, and the last imaginary term to thermal fluctuations.  
Since this is a quadratic functional of $X(t)$, $Y(t)$, $\xi(t)$, and $\eta(t)$, then the propagator ( \ref{J_ba}) can be calculated exactly and can be written as
\begin{equation}
  J(X_b,Y_b,\xi_b,\eta_b,t_b|X_a,Y_a,\xi_a,\eta_a,t_a) 
  \!=\! N\!(t) \exp\!\left\{\!\frac{i\bar{S}}{\hbar} \!\right\},~~
  \label{J0_ba}
\end{equation}
where $N(t)$ is a normalization factor and $\bar{S}$ is the effective action $S$ computed along the trajectories which make it an extremum,
\begin{eqnarray}
  \!\!\! \frac{1}{m}\frac{\delta S}{\delta\xi(t)} 
  &=& \ddot{X}(t)+\gamma \dot{X}(t)-\omega\dot{Y}(t)+\frac{2i\gamma}{\beta\hbar}\xi(t)=0,
  \label{X_eq}
  \\
  \!\!\! \frac{1}{m}\frac{\delta S}{\delta\eta(t)} 
  &=& \ddot{Y}(t)+\gamma \dot{Y}(t)+\omega\dot{X}(t)+\frac{2i\gamma}{\beta\hbar}\eta(t)=0,
  \label{Y_eq}
  \\
  \!\!\! \frac{1}{m}\frac{\delta S}{\delta X(t)} 
  &=& \ddot{\xi}(t)+\gamma \dot{\xi}(t)-\omega\dot{\eta}(t)=0,
  \label{xi_eq}
  \\
  \!\!\! \frac{1}{m}\frac{\delta S}{\delta Y(t)} 
  &=& \ddot{\eta}(t)+\gamma \dot{\eta}(t)+\omega\dot{\xi}(t)=0,
  \label{eta_eq}
\end{eqnarray}
subject to the boundary conditions $a$ and $b$.
The calculation is simplified by integrating by parts Eq.~(\ref{S_eff1}) and setting $\xi_b=\eta_b=0$ in view of the calculation of the probability density, which provides
\begin{eqnarray}
  \left( \bar{S} \right)_{\xi_b=\eta_b=0}
  &=& m \left[ \dot{X}_a \xi_a + \dot{Y}_a \eta_a
      - \frac{\omega}{2} \left( Y_a \xi_a - X_a \eta_a \right) \right]
  \nonumber \\
  &-& \frac{ i m \gamma }{ \beta \hbar } 
      \int_{t_a}^{t_b} \!\!\! dt {\left[ \xi^2(s) + \eta^2(s) \right]}_{\xi_b=\eta_b=0}.
\label{S_eff2}
\end{eqnarray}
Here only the solutions $\xi(t)$ and $\eta(t)$ of Eqs.~(\ref{xi_eq}) and (\ref{eta_eq}) with final conditions $\xi_b = \eta_b = 0$ are needed.
In analogy with the classical case, using the coordinates $Z = X +iY$ and $\zeta = \xi + i\eta$, Eqs.~(\ref{X_eq})--(\ref{eta_eq}) become 
\begin{eqnarray}
  \ddot{Z}(t)+\widetilde{\varGamma} \dot{Z}(t) = - \frac{2i\gamma}{\beta\hbar}\zeta(t) \ ,
  ~~~~~
  \ddot{\zeta}(t) - \widetilde{\varGamma}^* \dot{\zeta}(t) = 0 \ .
  \label{zeta_Zeta}
\end{eqnarray}
We skip the details that will be presented elsewhere and provide the result,
\begin{eqnarray}
  &&\left( \bar{S} \right)_{\xi_b=\eta_b=0}
      = m \Big\{ \left[ \mathcal{R}(t) \xi_a + \mathcal{I}(t) \eta_a \right] (X_b-X_a)
  \nonumber
  \\
  && ~~~~~ +\left[ \mathcal{R}(t) \eta_a - \mathcal{I}(t) \xi_a \right] (Y_b-Y_a)
  \nonumber
  \\
  && ~~~~~ +\frac{\omega}{2} \left(  X_a \eta_a - Y_a \xi_a \right)
    + \frac{ 2 i m \gamma }{ \beta \hbar } 
    \epsilon(t) \left( \xi_a^2 + \eta_a^2 \right) \Big\},~~~~~
\label{S_eff3}
\end{eqnarray}
where
\begin{eqnarray}
  &&\mathcal{R}(t) + i \mathcal{I}(t) = \frac{\widetilde{\varGamma}}{1-\exp[-\widetilde{\varGamma}t]} \,
  \label{RI}
  \\
  &&\epsilon(t) \!=\! 
  \frac{1}{M}
  \!\bigg[ t 
     \!+\! \frac{2\mathrm{e}^{-\gamma t}}{\varGamma}\cos(\omega t \!+\! \chi_\mathrm{H})
     \!-\! \frac{2\gamma}{\varGamma^2}
     \!+\! \frac{1 \!-\! \mathrm{e}^{-2\gamma t}}{2\gamma} 
  \bigg] \! ,
  \label{epsilon}
  \\
  && M(t)
  = 1 - 2\mathrm{e}^{-\gamma t}\cos(\omega t) + \mathrm{e}^{-2\gamma t}.
  \label{M}
\end{eqnarray}
%

% --------------------------------------------------------------------------------
\subsection{Probability density}
\label{Probability}
% --------------------------------------------------------------------------------

The effective propagator (\ref{J0_ba}), with the effective action (\ref{S_eff3}), and the initial conditions (\ref{Rho_a}) can be used in Eq.~(\ref{P_b}) to compute the probability density $P(X_b,Y_b,t_b)$.
The first two integrations give
\begin{eqnarray}
  P(X_b,Y_b,t_b) &=& f(t) \int d\xi_a d\eta_a  \exp\big( -A_0\xi_a^2 - B_0\eta_a^2
    \nonumber \\
                 &+& 2 C_0 \xi_a \eta_a + D_0 \xi_a + E_0\eta_a \big),
  \label{P_b2}
\end{eqnarray}
where
\begin{eqnarray}
  A_0(t) &=& \frac{m^2}{2\hbar^2} \left[ \sigma_x^2 \mathcal{R}(t)^2 + \sigma_y^2 \mathcal{I}_+(t)^2 +
        \frac{\hbar}{2m\sigma_x^2} + \frac{2\gamma\epsilon(t)}{m\beta} \right] \ ,
  \nonumber \\
  B_0(t) &=& \frac{m^2}{2\hbar^2} \left[ \sigma_y^2 \mathcal{R}(t)^2 + \sigma_x^2 \mathcal{I}_-(t)^2 +
        \frac{\hbar}{2m\sigma_y^2} + \frac{2\gamma\epsilon(t)}{m\beta} \right] \ ,
  \nonumber \\
  C_0(t) &=& \frac{m^2}{2\hbar^2} \left[ \sigma_x^2 \mathcal{I}_+(t)^2 - \sigma_x^2 \mathcal{I}_-(t)^2 +
        \frac{\hbar}{2m\sigma_y^2} + \frac{2\gamma\epsilon(t)}{m\beta} \right] \ ,
  \nonumber \\
  D_0(t) &=& \frac{im}{\hbar} \left[ \mathcal{R}(t) (X_b-x_0) - \mathcal{I}(t) (Y_ - y_0) - v_x \right] \ ,
  \nonumber \\
  E_0(t) &=& \frac{im}{\hbar} \left[ \mathcal{I}(t) (X_b-x_0) + \mathcal{R}(t) (Y_b - y_0) - v_y \right] \ ,
  \label{ABCDE}
\end{eqnarray}
where $\mathcal{I}_\pm(t) = \mathcal{I}(t) + \frac{1}{2} (\pm\alpha - 1) \omega$ and $\alpha$ is the gauge parameter given by Eq.~(\ref{alpha}).
The last integrations give
\begin{eqnarray}
  P(X_b,Y_b,t_b) &=& N(t) \exp\big( - A_1 X_b^2 - B_1 Y_b^2
    \nonumber \\
                 &+& 2C_1 X_b'Y_b'+ D_1 X_b'+ E_1 Y_b' \big),
  \label{P_b3}
\end{eqnarray}
where $N(t)$ is a suitable normalization factor, $X_b'$ and $Y_b'$
the coordinates relative to the initial mean position,
\begin{equation}
  X_b' = X_b - x_0, ~~~~ Y_b' = Y_b - y_0 \ ,
  \label{XY_prime}
\end{equation}
and the time-dependent coefficients are given by
\begin{eqnarray}
  A_1 &=& \frac{(m/2\hbar)^2}{A_0B_0-C_0^2} 
           \left[ B_0\mathcal{R}^2 + 2 C_0\mathcal{R}\mathcal{I} + A_0\mathcal{I}^2 \right] \! ,
  \nonumber \\
  B_1 &=& \frac{(m/2\hbar)^2}{A_0B_0-C_0^2} 
           \left[ B_0\mathcal{I}^2 - 2 C_0\mathcal{R}\mathcal{I} + A_0\mathcal{R}^2 \right] \! ,  
  \nonumber \\
  C_1 &=&  \frac{(m/2\hbar)^2}{A_0B_0-C_0^2} 
           \left[ (A_0-B_0)\mathcal{R}\mathcal{I} + C_0(\mathcal{R}^2-\mathcal{I}^2)\right] \! ,
  \nonumber \\
  D_1 &=&  \frac{2(m/2\hbar)^2}{A_0B_0-C_0^2} 
           \left[ (B_0\mathcal{R}+C_0\mathcal{I}) v_x + (C_0\mathcal{R}+A_0\mathcal{I}) v_y \right] \! ,
  \nonumber \\
  E_1 &=&   \!\frac{2(m/2\hbar)^2}{A_0B_0 \!-\! C_0^2} 
           \left[ (C_0\mathcal{R} \!-\! B_0\mathcal{I}) v_x + (A_0\mathcal{R} \!-\! C_0\mathcal{I}) v_y \right] \! .
  \label{ABCDE_1}
\end{eqnarray}
%

% --------------------------------------------------------------------------------
\subsection{Mean trajectory}
\label{trajectory}
% --------------------------------------------------------------------------------

The mean coordinates at the generic time $t_b$ are
\begin{eqnarray}
 \!\! \left\langle X \right\rangle_t \!&=&\! \!\int \! dX_b dY_b \,X_b \, P(X_b,Y_b,t_b)
              \!=\! x_0 \!+\! \frac{\mathcal{R} v_x + \mathcal{I} v_y}{2(\mathcal{R}^2 + \mathcal{I}^2)} \!,
  \nonumber \\
 \!\! \left\langle Y \right\rangle_t \!&=&\! \!\int \! dX_b dY_b \,Y_b \, P(X_b,Y_b,t_b)
              \!=\! y_0 \!+\! \frac{\mathcal{R} v_y - \mathcal{I} v_x}{2(\mathcal{R}^2 + \mathcal{I}^2)} \!.
  \label{XY_b}
\end{eqnarray}
By using the explicit formulas for $\mathcal{R}(t)$ and $\mathcal{I}(t)$ given by Eqs.~(\ref{RI}), these equations reduce just to the classical solutions (\ref{xy-E0}) and (\ref{xcyc}), with $t$ replaced by $t_b-t_a$ and $x_0$, $y_0$, $v_x, v_y$ by the corresponding quantum mean values at $t=t_a$.
%, as expected also from Ehrenfest's theorem \cite{Schiff1968a}.

% --------------------------------------------------------------------------------
\subsection{Position uncertainties}
\label{dr2}
% --------------------------------------------------------------------------------

The second central moments of the coordinates, i.e., 
\begin{eqnarray}
  \left\langle \Delta X^2 \right\rangle_t 
     &=& \int dX_b dY_b \, [X_b - \left\langle X \right\rangle_t]^2 P(X_b,Y_b,t_b)
     \nonumber \\
     &=& 2\left(\frac{\hbar}{m}\right)^2
     \frac{B_0\mathcal{I}^2-2C_0\mathcal{R}\mathcal{I}+A_0\mathcal{R}^2}{(\mathcal{R}^2 + \mathcal{I}^2)^2},
     \nonumber \\
  \left\langle \Delta Y^2 \right\rangle_t &=& \int dX_b dY_b \, [Y_b - \left\langle Y \right\rangle_t]^2 P(X_b,Y_b,t_b)
     \nonumber \\
      &=& 2\left(\frac{\hbar}{m}\right)^2
      \frac{B_0\mathcal{R}^2+2C_0\mathcal{R}\mathcal{I}+A_0\mathcal{I}^2}{(\mathcal{R}^2 + \mathcal{I}^2)^2},
      \nonumber \\
  \left\langle \Delta \! X \Delta \! Y \right\rangle_{\!t} 
      &=& \!\! \int \!\! dX_b dY_b [X_b \!-\! \left\langle X \right\rangle_t][Y_b \!-\! \left\langle Y \right\rangle_t] P(X_b,\!Y_b,\!t_b)
      \nonumber \\
      &=& \!2\!\left(\!\frac{\hbar}{m}\!\right)^{\! 2}
                \frac{(B_0 \!-\! A_0)\mathcal{R}^2 \!+\! C_0(\mathcal{I}^2 \!-\! A_0\mathcal{R}^2)}{(\mathcal{R}^2 \!+\! \mathcal{I}^2)^2}.
  \label{DXY}
\end{eqnarray}
provide estimates of the quantum uncertainties on the particle position.
With the help of Eqs. (\ref{RI}), one obtains
\begin{eqnarray}
  &&\left\langle \Delta X^2 \right\rangle_t 
      = \frac{\Delta r^2}{2}
      + \left(\frac{\hbar g_I}{m\sigma_y}\right)^2
      + \left(\frac{\hbar g_R}{m\sigma_x}\right)^2
      \nonumber \\
  && ~~~ + \sigma_x^2 [1-(1+\alpha)\omega g_I]^2 
      + \sigma_y^2 [(1-\alpha)\omega g_R]^2 \ ,~~~
      \nonumber
      \\
  &&\left\langle \Delta Y^2 \right\rangle_t 
      = \frac{\Delta r^2}{2}
      + \left(\frac{\hbar g_I}{m\sigma_x}\right)^2
      + \left(\frac{\hbar g_R}{m\sigma_y}\right)^2
      \nonumber\\
  && ~~~ + \sigma_y^2 [1-(1-\alpha)\omega g_I]^2 
       +  \sigma_x^2 [(1+\alpha)\omega g_R]^2 \ ,~~~
      \nonumber\\
  &&\left\langle \Delta X \Delta Y \right\rangle_t 
      = \left(\frac{\hbar g_R}{m}\right)^2 \left( \frac{1}{\sigma_y^2}-\frac{1}{\sigma_x^2}\right)
      \nonumber \\
  &&+ 4 (g_I^2 \!-\! g_R^2)\mathcal{R}  \bigg[
         \!\!\left(\mathcal{R} \!-\! \mathcal{I} \!+\! \frac{\omega}{2}\right) \! (\sigma_x^2 \!-\! \sigma_y^2)
         \!+\! \frac{\alpha\omega(\sigma_x^2 \!+\! \sigma_y^2)}{2}
      \bigg]\!,~~~
  \label{DXY1}
\end{eqnarray}
where $g_{\rm I}(t) = (M\mathcal{I})/(2|\varGamma|^2)$ and $g_{\rm R}(t) = (M\mathcal{R})(2|\varGamma|^2)$.

In the asymptotic limit $\gamma t \gg 1$, one obtains
\begin{eqnarray}
  \left\langle\! \Delta X^2 \!\right\rangle_t \!\to\!  2 D_B t,
      \nonumber\\
  \left\langle\! \Delta Y^2 \!\right\rangle_t \!\to\!  2 D_B t ,
      \nonumber\\
  \left\langle \Delta X \Delta Y \right\rangle_t \!\to\! \mathrm{const} .
  \label{DXY2}
\end{eqnarray}
%

% --------------------------------------------------------------------------------
\subsection{Evolution of the wave packet}
\label{wave_packet}
% --------------------------------------------------------------------------------

The Gaussian probability wave packet obtained above may not have an obvious interpretation, apart in the asymptotic limit.
A simple way to an effective visualization is the method of the boundary surface \cite{Atkins1986a}, in which a space-dependent probability density $P$ is represented through an iso-probability surface in 3D, $P = \mathrm{const}$, enclosing an assigned fraction of the total probability, e.g. 90\%.
In the present 2D case, using Eq.~(\ref{P_b3}), one can define an iso-probability curve in one of the following ways,
\begin{eqnarray}
&&\!\!P(X,Y,t) \propto \exp[-g(X,Y)] = \kappa,~~
\\
&&\!\!g(X,Y) \!=\!\! A_1 X^2 \!\!+\! B_1 Y^2 \!\!-\! 2C_1 XY \!\!-\! D_1 X \!\!-\! E_1 Y  \!\!\!=\! \kappa',~~
\label{Pg}
\end{eqnarray}
where $\kappa,\kappa'$ are suitable constants.
The second equation defines an ellipse in the $X$-$Y$-plane with time-varying center and axis orientation.

To visualize it, one can first move to a reference frame $(X_1,Y_1)$ where the ellipse center is at rest, by introducing the coordinates relative to the ellipse center, i.e., to the average position and therefore to the classical solution,
\begin{eqnarray}
      X_1 = X - \left\langle X \right\rangle_t \ ,
      ~~~~~
      Y_1 = Y - \left\langle Y \right\rangle_t \ .
  \label{XY1}
\end{eqnarray}
The new boundary curve,
\begin{equation}
  A_1 X_1^2 + B_1 Y_1^2 - 2C_1 X_1 Y_1 = \mathrm{const} \ ,
  \label{bs1}
\end{equation}
defines an ellipse with center in the origin.

The ellipse is not in normal form yet: the presence of the mixed term $\propto X_1 Y_1$ with the time-dependent coefficient $C_1$ implies a rotation in time of the principal axes of the ellipse.
 One can move to a reference frame $X_2$-$Y_2$ rotating with the ellipse, defined by
\begin{eqnarray}
      X_2 = X_1 \cos\theta + Y_1 \sin\theta,
      ~
      Y_2 = -X_1 \sin\theta + Y_1 \cos\theta,~~~~~
  \label{XY_2}
\end{eqnarray}
where $\theta(t)$ is a suitable function of time.
The new boundary curve is defined by
\begin{eqnarray}
  \label{bs2}
  &&A_2 X_2^2 + B_2 Y_2^2 - 2C_2 X_2 Y_2 = \mathrm{const} \ ,
      \nonumber \\
  &&A_2 = \frac{1}{2}\left[ A_1+B_1+(A_1-B_1)\cos(2\theta)-2C_1\sin(2\theta) \right] \ ,
      \nonumber \\
  &&B_2 = \frac{1}{2}\left[ A_1+B_1-(A_1-B_1)\cos(2\theta)+2C_1\sin(2\theta) \right] \ ,
      \nonumber \\
  &&C_2 = \frac{1}{2}\left[ (A_1-B_1)\sin(2\theta)+2C_1\cos(2\theta) \right] \ .
\end{eqnarray}
The mixed term (and the rotational motion) is removed by setting $C_2=0$, which defines $\theta(t)$ as
\begin{eqnarray}
  \label{theta}
  \tan\theta &=& \frac{2C_1}{B_1-A_1} \equiv -\tan(\theta_1+\theta_2) \ ,
      \\
  \tan\theta_1 &=& \frac{2\mathcal{R}\mathcal{I}}{\mathcal{R}^2-\mathcal{I}^2} \ ,
      \nonumber \\
  \tan\theta_2 &=& \frac{2\mathcal{R}(\sigma_y^2 I_+ -\sigma_x^2 I_-)}
                     {\sigma_y^2 I_+^2
                      \!-\!
                      \sigma_x^2 I_-^2
                      \!+\!
                      (\sigma_x^2-\sigma_y^2)\mathcal{R}^2
                      \!+\!
                     \left(\! \frac{\hbar}{2m} \!\right)^2 \!\! (\sigma_x^{-2}-\sigma_y^{-2})} \ .
    \nonumber 
\end{eqnarray}
In the new variables, the boundary curve reads
\begin{equation}
  A_2(t) X_2^2 + B_2(t) Y_2^2 = \mathrm{const} \ ,
  \label{bs3}
\end{equation}
which represents an ellipse with $X_2$ and $Y_2$ principal axis proportional to $1/\sqrt{A_2}$ and $1/\sqrt{B_2}$, respectively.

In conclusion, one can give an intuitive interpretation of the wave packet evolution in terms of a superposition of (a) a translational motion defined by the classical solution, (b) a rotational motion defined with angular velocity $\dot{\theta}(t)$, where $\theta$ is defined by Eq. (\ref{theta}), and (3) a spreading process defined by the second moments $1/A_2^2$ and $1/B_2^2$ of the ellipse recast in normal form.
Detailed illustrations will be presented elsewhere.

% ==============================================================================
\section{Conclusions}
\label{conclusion}
% ==============================================================================

The oscillator model of quantum dissipative systems has been used as an effective tool for the visualization of the dynamics of a quantum Brownian particle in configuration space.
Due to the quadratic nature of the system Lagrangian, the center of mass of the probability density obtained from the solution of the problem moves like a classical particle and the shape of an initially Gaussian probability density remains Gaussian.
As for the particle uncertainty on position, the width of the probability increases with time depending on both quantum and thermal fluctuations.
Interestingly, the probability wave packet in configuration space can be visualized as if it rotates around the center of mass with an angular velocity depending on the system parameters.
Further research work is needed for an understanding of this effect.
Furthermore, it would be worth to study the analogous problem in the presence of colored noise.

\vspace{0.5cm}

% ==============================================================================
\section*{Acknowledgements}
% ==============================================================================

M.P. acknowledges support from the Estonian Science Foundation Grant no. 9462
and by institutional research funding IUT (IUT-39) of the Estonian Ministry of Education and Research.

\bibliography{local_QBM-magnetic}

\begin{thebibliography}{10}

\bibitem{FQMT15}
Presented at {FQMT15}--{F}rontiers of {Q}uantum and {M}esoscopic
  {T}hermodynamics, {J}uly 27-{A}ugust 1, 2015, {P}rague, {C}zech
  {R}epublic~~(\url{http://fqmt.fzu.cz/15/}).

\bibitem{Ashcroft1976a}
N.~W. Ashcroft and N.~D. Mermin.
\newblock {\em Solid State Physics}.
\newblock Saunders College, Philadelphia, 1976.

\bibitem{Atkins1986a}
P.~W. Atkins.
\newblock {\em Physical Chemistry}.
\newblock Oxford University Press, Oxford, third edition, 1986.

\bibitem{Caldeira1983a}
A.~O. Caldeira and A.~J. Leggett.
\newblock {Path integral approach to quantum Brownian motion}.
\newblock {\em Physica A}, 121:587, 1983.

\bibitem{Czopnik2001a}
R.~Czopnik and P.~Garbaczewski.
\newblock Brownian motion in a magnetic field.
\newblock {\em Phys. Rev. E}, 63:021105, 2001.

\bibitem{Czopnik2001b}
R.~Czopnik and P.~Garbaczewski.
\newblock Charged {B}rownian particle in a magnetic field.
\newblock {\em Acta Physica Pol. B}, 32(5):1437, 2001.

\bibitem{Das1981a}
A.~K. Das.
\newblock {{B}rownian motion in a magnetic field: A model for a semi-quantum
  stoschastic process}.
\newblock {\em Z. Phys. B}, 40:353, 1981.

\bibitem{Das1982a}
A.~K. Das.
\newblock {{B}rownian motion in a quantizing magnetic field}.
\newblock {\em Physica A}, 110:489, 1982.

\bibitem{Dattagupta2014a}
S.~Dattagupta.
\newblock {\em Diffusion. Formalism and Applications}.
\newblock CRC Press Taylor \& Francis, Boca Raton, 2014.

\bibitem{Dattagupta2010a}
S.~Dattagupta, J.~Kumar, S.~Sinha, and P.~A. Sreeram.
\newblock Dissipative quantum systems and the heat capacity.
\newblock {\em Phys. Rev. E}, 81:031136, 2010.

\bibitem{Dattagupta1996a}
S.~Dattagupta and J.~Singh.
\newblock Stochastic motion of a charged particle in a magnetic field: Ii
  quantum {B}rownian treatment.
\newblock {\em Pramana J. Phys.}, 47:211, 1996.

\bibitem{Dodin2005a}
I.Y. Dodin and N.J. Fisch.
\newblock Ponderomotive ratchet in a uniform magnetic field.
\newblock {\em Phys. Rev. E}, 72:046602, 2005.

\bibitem{Feynman1965a}
R.~P. Feynman and A.~R. Hibbs.
\newblock {\em Quantum Mechanics and Path Integrals}.
\newblock McGraw-Hill, N.Y., 1965.

\bibitem{Feynman1963a}
R.~P. Feynman and F.~L. Vernon.
\newblock {The theory of a general quantum system interacting with a linear
  dissipative system}.
\newblock {\em Ann. Phys. (N.Y.)}, 24:118, 1963.

\bibitem{Friz2015a}
P.~Friz, P.~Gassiat, and T.~Lyons.
\newblock Physical brownian motion in a magnetic field as a rough path.
\newblock {\em Trans. Am. Math. Soc.}, 367(11):7939--7955, 2015.

\bibitem{Furuse1970a}
H.~Furuse.
\newblock Influence of magnetic field on the {B}rownian motion of charged
  particle.
\newblock {\em J. Phys. Soc. Japan}, 28(3):559, 1970.

\bibitem{Hou2009a}
L.~J. Hou, Z.~L. Mi\v{s}kovi{\'c}, A.~Piel, and P.~K. Shukla.
\newblock Brownian dynamics of charged particles in a constant magnetic field.
\newblock {\em Physics of Plasmas}, 16(5):053705, 2009.

\bibitem{Jayannavar1981a}
A.~M. Jayannavar and N.~Kumar.
\newblock Orbital diamagnetism of a charged {B}rownian particle undergoing a
  birth-death process.
\newblock {\em J. Phys. A: Math. Gen.}, 14:1399, 1981.

\bibitem{Jimenez2006a}
J.~I. Jim{\'e}nez-Aquino and M.~Romero-Bastida.
\newblock Fokker-planck-kramers equation for a brownian gas in a magnetic
  field.
\newblock {\em Phys. Rev. E}, 74:041117, 2006.

\bibitem{Jimenez2007a}
J.~I. Jim{\'e}nez-Aquino and M.~Romero-Bastida.
\newblock Fokker-planck-kramers equations of a heavy ion in presence of
  external fields.
\newblock {\em Phys. Rev. E}, 76:021106, 2007.

\bibitem{Jimenez2008c}
J.~I. Jim{\'e}nez-Aquino, M.~Romero-Bastida, and A.C. P{\'e}rez-Guerrero
  Noyola.
\newblock Brownian motion in a magnetic field and in the presence of additional
  external forces.
\newblock {\em Revista Mexicana de F{\'i}sica E}, 54(10):81--86, 2008.

\bibitem{Karmeshu1973a}
Karmeshu.
\newblock Velocity fluctuations of charged particles in the presence of
  magnetic field.
\newblock {\em J. Phys. Soc. Japan}, 34:1467, 1973.

\bibitem{Karmeshu1974a}
Karmeshu.
\newblock Brownian motion of charged particles in a magnetic field.
\newblock {\em Phys. Fluids}, 17:1828, 1974.

\bibitem{Kursunoglu1962a}
B.~Kur\c{s}uno\u{g}lu.
\newblock Brownian motion in a magnetic field.
\newblock {\em Ann. Phys.}, 17:259, 1962.

\bibitem{Kursunoglu1963a}
B.~Kur\c{s}uno\u{g}lu.
\newblock Brownian motion in a magnetic field.
\newblock {\em Phys. Rev.}, 132:211, 1963.

\bibitem{Lagos2011a}
R.~E. Lagos and T.~P. Sim{\~o}es.
\newblock Charged brownian particles: Kramers and smoluchowski equations and
  the hydrothermodynamical picture.
\newblock {\em Physica A}, 390(9):1591--1601, 2011.

\bibitem{Landau2}
L.~D. Landau and E.~M. Lifchitz.
\newblock {\em The Classical Theory of Fields}.
\newblock Course of Theoretical physics. Elsevier, Amsterdam, 1975.

\bibitem{Landau3}
L.~D. Landau and E.~M. Lifsits.
\newblock {\em Quantum Mechanics}, volume~3 of {\em Course of Theoretical
  physics}.
\newblock Pergamon Press, Oxford, 1965.

\bibitem{Lemons1999a}
D.S. Lemons and D.L. Kaufman.
\newblock Brownian motion of a charged particle in a magnetic field.
\newblock {\em IEEE Trans. on Plasma Sci.}, 27(5):1288--1296, 1999.

\bibitem{Li1990a}
X.~L. Li, G.~W. Ford, and R.~F. O'Connel.
\newblock Magnetic-field effects on the motion of a charged particle in a heat
  bath.
\newblock {\em Phys. Rev. A}, 41:5287, 1990.

\bibitem{Liboff1966a}
R.~L. Liboff.
\newblock Brownian motion of charged particles in crossed. electric and
  magnetic fields.
\newblock {\em Phys. Rev.}, 141:222, 1966.

\bibitem{Marathe1989a}
Y.~Marathe.
\newblock Dissipative quantum dynamics of a charged particle in a magentic
  field.
\newblock {\em Phys. Rev. A}, 39:5927, 1989.

\bibitem{Mitra2010a}
A.~N. Mitra.
\newblock Can environmental decoherence be reversed for an open quantum system
  in a magnetic field?
\newblock {\em arXiv:1007.0168}, 2010.

\bibitem{Paraan2008a}
F.~N.~C. Paraan, M.~P. Solon, and J.~P. Esguerra.
\newblock Brownian motion of a charged particle driven internally by correlated
  noise.
\newblock {\em Phys. Rev. E}, 77:022101, 2008.

\bibitem{Patriarca1996a}
M.~Patriarca.
\newblock {Statistical correlations in the oscillator model of quantum
  {B}rownian motion}.
\newblock {\em Il Nuovo Cimento B}, 111:61, 1996.

\bibitem{Simoes2005a}
T.~P. Sim{\~o}es and R.~E. Lagos.
\newblock Kramers equation for a charged brownian particle: The exact solution.
\newblock {\em Physica A}, 355:274--282, 2005.

\bibitem{Singh1996a}
J.~Singh and S.~Dattagupta.
\newblock Stochastic motion of a charged particle in a magnetic field: I
  classical treatment.
\newblock {\em Pramana J. Phys.}, 47:199, 1996.

\bibitem{Taylor1961a}
J.~B. Taylor.
\newblock {Diffusion of plasma across a magnetic field}.
\newblock {\em Phys. Rev. Lett.}, 6:262, 1961.

\bibitem{Voropajeva2008a}
N.~Voropajeva and T.~{\"O}rd.
\newblock Correlation in the velocity of a {B}rownian particle induced by
  frictional anisotropy and magnetic field.
\newblock {\em Phys. Lett. A}, 372:2167, 2008.

\bibitem{Wang1945a}
M.~C. Wang and G.~E. Uhlenbeck.
\newblock {On the theory of {B}rownian motion {\rm II}}.
\newblock {\em Rev. Mod. Phys.}, 17:323, 1945.

\bibitem{Williamson1968a}
J.~H. Williamson.
\newblock {{B}rownian motion of electrons}.
\newblock {\em J. Phys. A}, 1:629, 1968.

\bibitem{Xiang1993a}
N.~Xiang.
\newblock Stochastic motion of charged particles in a magnetic field.
\newblock {\em Phys. Rev. E}, 48:1590, 1993.

\end{thebibliography}
\bibliographystyle{plain}

\end{document}